\documentclass{aa}  
\usepackage{natbib,twoopt}
\usepackage[breaklinks=true]{hyperref} 
\bibpunct{(}{)}{;}{a}{}{,}             
\makeatletter
  \newcommandtwoopt{\citeads}[3][][]{\href{http://adsabs.harvard.edu/abs/#3}%
    {\def\hyper@linkstart##1##2{}%
     \let\hyper@linkend\@empty\citealp[#1][#2]{#3}}}
  \newcommandtwoopt{\citepads}[3][][]{\href{http://adsabs.harvard.edu/abs/#3}%
    {\def\hyper@linkstart##1##2{}%
     \let\hyper@linkend\@empty\citep[#1][#2]{#3}}}
  \newcommandtwoopt{\citetads}[3][][]{\href{http://adsabs.harvard.edu/abs/#3}%
    {\def\hyper@linkstart##1##2{}%
     \let\hyper@linkend\@empty\citet[#1][#2]{#3}}}
  \newcommandtwoopt{\citeyearads}[3][][]%
    {\href{http://adsabs.harvard.edu/abs/#3}
    {\def\hyper@linkstart##1##2{}%
     \let\hyper@linkend\@empty\citeyear[#1][#2]{#3}}}
\makeatother

\usepackage{graphicx}
\usepackage{txfonts}

\begin{document} 

\title{A new low-luminosity globular cluster discovered in the Milky Way with the VVVX survey}
   \author{E. R. Garro \inst{1} 
          \and
          D. Minniti\inst{1,2}
           \and
             M. Gómez \inst{2}
               \and
             J.~G.~Fernández-Trincado\inst{3}
             \and
             J.  Alonso-García\inst{4,5}  
             \and
             M. Hempel \inst{1,6}
             \and
             R. Zelada Bacigalupo \inst{7}
          }
   \institute{Departamento de Ciencias Físicas, Facultad de Ciencias Exactas, Universidad Andres Bello, Fernández Concha 700, Las Condes, Santiago, Chile
   \and
 Vatican Observatory, Vatican City State, V-00120, Italy
 \and
 Instituto de Astronom\'ia, Universidad Cat\'olica del Norte, Av. Angamos 0610, Antofagasta, Chile
 \and
 Centro de Astronomía (CITEVA), Universidad de Antofagasta, Av. Angamos 601, Antofagasta, Chile
 \and
 Millennium Institute of Astrophysics , Nuncio Monse\~nor Sotero Sanz 100, Of. 104, Providencia, Santiago, Chile
 \and
 Max-Planck Institute for Astronomy, Koenigstuhl 17, 69177 Heidelberg,
Germany
 \and
  North Optics Instrumentos Científicos, La Serena, Chile
}
  \date{Received: February 17,  2022; Accepted: May 2, 2022}
 
\abstract
   {
The VISTA Variables in the Via Láctea Extended Survey (VVVX) allows us to probe previously unexplored regions of the inner Milky Way (MW), especially those that are affected by stellar crowding and strong extinction.
   }
   {
Our long-term goal is to identify new star clusters and investigate them to reveal their true nature. In particular, we are looking for new candidate globular clusters (GCs) located in the Galactic bulge and disk, with the aim of completing the census of the MW GC system.
   }
   {
We searched and characterised new GCs, using a combination of the near-infrared (IR) VVVX survey and Two Micron All Sky survey (2MASS) datasets, and the optical Gaia Early Data Release 3 (EDR3) photometry and its precise proper motions (PMs). 
   }
   {
We report the discovery of a new Galactic GC, named Garro 02, situated at RA~=~18:05:51.1, Dec~=~-17:42:02 and  $l =12^{\circ}.042$,  $b=+1^{\circ}.656$.  Performing a PM-decontamination procedure, we built a final catalogue with all cluster members, on which we performed a photometric analysis. We calculated a reddening of $E(J-K_s) = 1.07 \pm 0.06$ mag and extinction of $A_{Ks} = 0.79\pm 0.04$ mag in the near-IR; while $E(BP-RP)=2.40\pm 0.01$ mag  and $A_{G}= 4.80\pm 0.02$ mag in optical passbands.  Its heliocentric distance is $D=5.6\pm0.8 $ kpc, which places Garro 02 at a Galactocentric distance of $R_{G}=2.9$ kpc and $Z=0.006$ kpc above the Galactic plane. We also estimated the metallicity and age by comparison with known GCs and by fitting PARSEC isochrones, finding [Fe/H]~$=-1.30 \pm 0.2$ dex and age $= 12\pm 2$ Gyr. We derived the mean cluster PM of $(\mu_{\alpha}^{\ast}, \mu_{\delta}) = (-6.07\pm 0.62 , -6.15\pm 0.75)$ mas yr$^{-1}$.  We calculated the cluster luminosity in the near-IR of $M_{Ks}=-7.52\pm 1.23$ mag, which is equivalent to $M_{V}=-5.44$ mag. The core and tidal radii from the radial density profile are $r_c = 1.25\pm 0.27$ arcmin (2.07 pc) and $r_{t} = 7.13\pm 3.83$ arcmin (11.82 pc), respectively.
   }
  {
We confirm Garro 02 as a new genuine Galactic GC, located in the MW bulge. It is a low-luminosity, metal-poor, and old GC, and it is a lucky survivor of the strong dynamical processes that occurred during the MW's entire life.
   }
  
   \maketitle
    
\section{Introduction}
\label{Introduction}
Recent studies (e.g. \citealt{Baumgardt2019}) have demonstrated that star clusters with lower masses and lower luminosities than those of known Galactic globular clusters (GCs) may be ubiquitous. We expect that these objects have been dissolved due to past strong dynamical processes, such as tidal shocks, evaporations,  and galaxy mergers (e.g. \citealt{Kruijssen2011}), but also they may be hidden in the Galactic bulge and disk \citep{Minniti2017,Garro2020}. Therefore, their identification seems to be very complicated. However, in the last decade, many new GC candidates have been discovered across the Milky Way (MW), especially in the bulge regions where the interstellar dust strongly hides the baryonic matter over the sky (e.g. \citealt{Gonzalez2012,Barbuy2018}) and the stellar crowding is substantially high.  Both the VISTA variables in the Vía Láctea (VVV; \citealt{Minniti2010,Saito2012}) and its eXtension (VVVX; \citealt{Minniti2018}) have substantially increased the number of
stellar cluster candidates in the MW. These studies use either
visual inspection or photometric analysis to search for henceforth
unknown stellar clusters and they have been extremely successful in that
respect (e.g. \citealt{Bica2019,Minniti2021_8SGR}).\\
Although the number of star cluster candidates has increased substantially,  more work has to be done to understand their nature,  for example to classify them as GCs (as well as open clusters or star associations). The Gaia mission (e.g. \citealt{CantatGaudin2019,CastroGinard2020}) is extremely helpful in that endeavour and has in fact confirmed many of the recently discovered star clusters as GCs,  some examples include the following: Garro 01 \citep{Garro2020}, FSR 1758 \citep{Barba2019, Villanova2019,Romero-Colmenares2021}, Patchick 99 \citep{Garro2021_P99}, Patchick 125 (\citealt{FernandezTrincado2021,ERG:submitted}, named also Gran 3 by \citealt{Gran2022}), and VVV-GC160 \citep{Minniti2021_VVV160,Garro2021_19GCs}; another nine GCs confirmed by \cite{Garro2021_19GCs},  FSR 19 and FSR 25 \citep{Obasi2021},  and Minni48 \citep{Minniti2021_M48}; and another five GCs confirmed in \cite{ERG:submitted}.  
These studies as well as other ones are fundamental since they can contribute, from a broad point of view, for one to understand the formation and evolution of the MW GC system.\\

In this paper, we focus on the discovery and confirmation of a new GC candidate in the MW (see Section \ref{Discovery}), which we named Garro 02. In Section \ref{datasets}, the datasets used to perform our analysis are presented.  The decontamination procedure and the selection of star cluster members are explained in Section \ref{Decontproc}.  In Section \ref{Parameters}, we derive its main physical parameters, such as reddening, extinction, mean cluster proper motions (PMs),  distance modulus, heliocentric and Galactocentric distances, age,  metallicity,  total luminosity, and finally structural parameters. A final summary and conclusions are given in Section \ref{Conclusions}.

\section{Discovery of a new GC candidate}
\label{Discovery}
\begin{figure*}
\centering
\includegraphics[width=6cm, height=5.5cm]{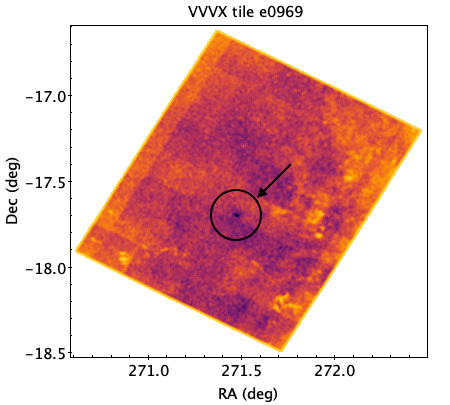} 
\includegraphics[width=6cm, height=5.5cm]{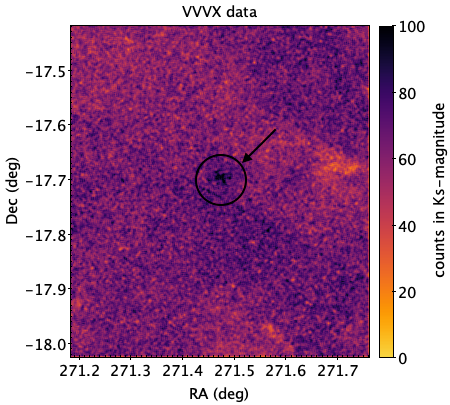} 
\includegraphics[width=6cm, height=5.5cm]{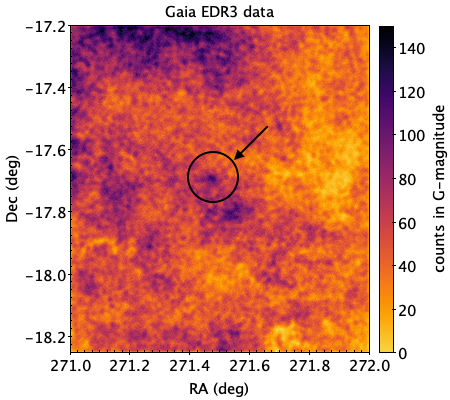} 
\caption{Density maps to search for the cluster as an over-density.  \textit{Left panel: }VVVX density map for the entire tile $e0969$ in equatorial coordinates.  The broad stripes seen in this image of the whole tile is the residual of the combination of the different pawprints. \textit{Middle and right panels:} Density maps of the zoomed region around Garro 02 using near-IR VVVX data and optical Gaia EDR3 data, respectively. We added a black circle and arrow in each panel in order to highlight the over-density, which indicates the presence of the cluster.  }
\label{fig:tile}
\end{figure*}

We report the discovery and confirmation of Garro 02,  a new GC located in the Galactic bulge at equatorial coordinates RA~=~18:05:51.1 and Dec~=~-17:42:02 and Galactic coordinates l = 12.042$^{\circ}$ and b = +1.656$^{\circ}$.  Following the same procedure described in \cite{Minniti2017}, we found this cluster using the VVVX database as a clear over-density of red giant branch stars (RGBs), above the background, in the tile $e0969$, as shown in Fig. \ref{fig:tile}.  We also observed this cluster in the optical wavelengths, using the Gaia EDR3 dataset.  Also in this case, we can see a clear excess of stars above the background, as the Gaia EDR3 source density map (Fig. \ref{fig:tile}) is clearly shown in a nearly circular region.\\

\begin{figure*}
\centering
\includegraphics[width=10cm, height=7cm]{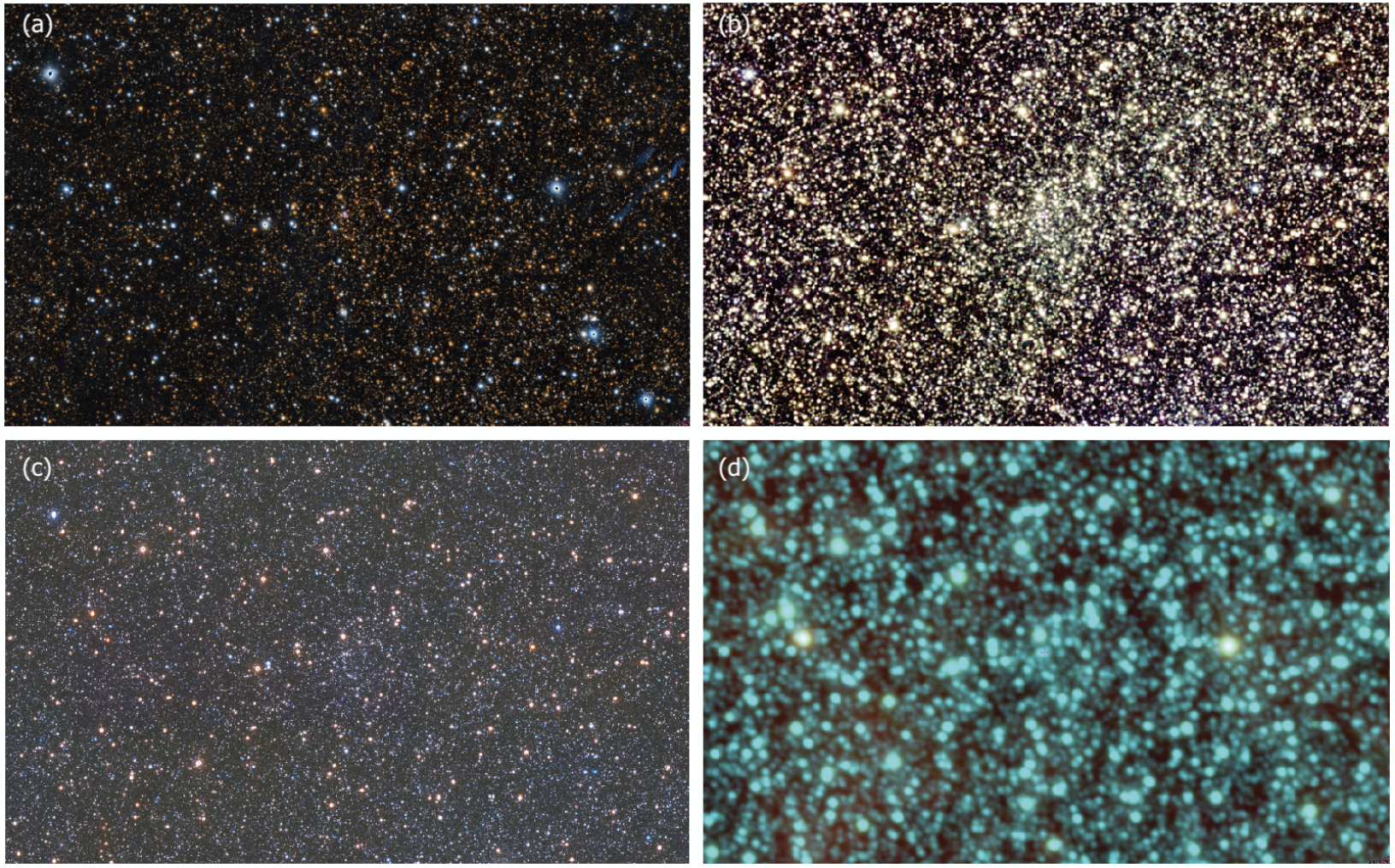} 
\caption{Pan-STARRS (a), 2MASS (b), VVVX (c), and WISE (d) images centred on Garro 02. These fields of view are about $21’\times 14’$, oriented along Galactic coordinates, with the longitude increasing to the left and the latitude increasing to the top. We show the Pan-STARRS and WISE images for comparison just to illustrate that the cluster is present at these wavelengths. The over-density is certainly better seen in the near-IR images (which is what we study in this paper), due to extinction and incompleteness of RGB stars in the other bands.}
\label{fig:finderchart}
\end{figure*}

However,  as demonstrated by previous works, such as \cite{Gran2019}, \cite{Palma2019}, and \cite{Minniti2021c}, not all over-densities are real clusters, but they may be simply a grouping of stars or statistical fluctuations of the projected stellar density in the plane of the sky.  Therefore, we performed four independent tests to confirm Garro 02 as a genuine GC.  First, we visually analysed this cluster at different wavelengths, using the Panoramic Survey Telescope and Rapid Response System (Pan-STARRS, \citealt{Chambers2016}), the Two Micron All Sky Survey (2MASS, \citealt{Skrutskie2006}), and the Wide-field Infrared Survey Explorer (WISE, \citealt{Wright2010}) images of the sky regions centred on Garro 02 as shown in Fig.  \ref{fig:finderchart}. Both the near-infrared (IR) 2MASS and mid-IR WISE surveys are sensitive to the clusters' bright RGBs; especially in the 2MASS image, a clear excess of RGB stars is visible at the Garro 02 coordinates, suggesting the presence of a cluster. 

\begin{figure*}
\centering
\includegraphics[width=16cm, height=5.5cm]{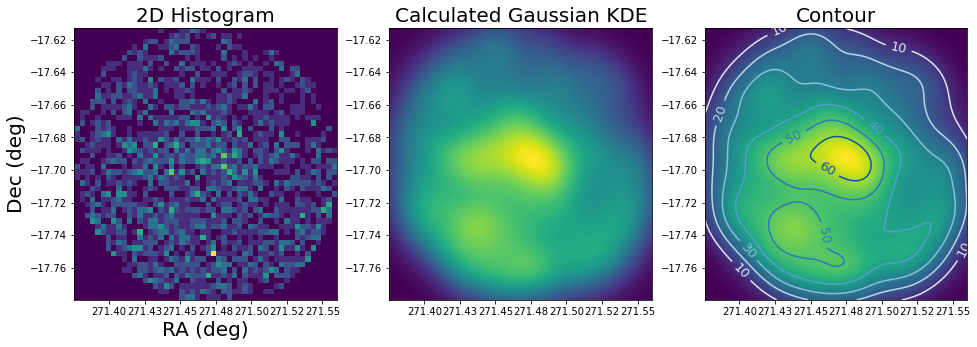} 
\includegraphics[width=16cm, height=5.5cm]{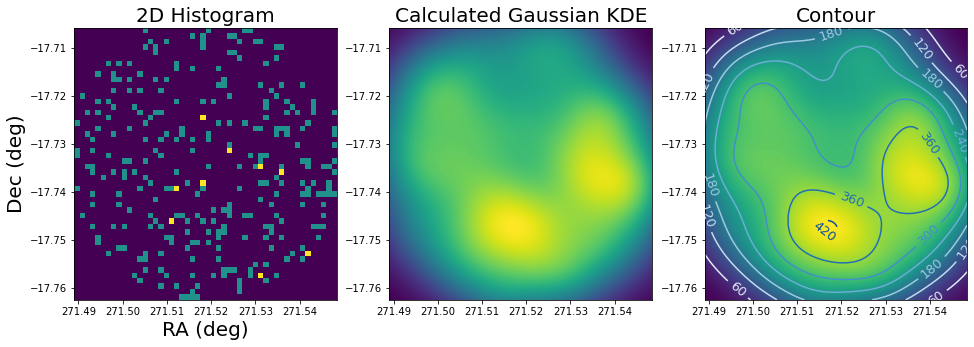} 
\caption{Spatial distribution for Garro 02 selected at $ r\lesssim 4'$ (on the top).  We also compare the cluster spatial distribution with its relative field (on the bottom) selected at $ 5' \lesssim r\lesssim 8'$ from the cluster centre.  We used the KDE technique in order to better distinguish the over-densities (yellow and green colours) from lower densities (blue colours). All samples were cleaned by nearby stars. These panels show the following from left to right: the 2D histograms and the calculated Gaussian KDE on the which we overplotted iso-density contours. }
\label{fig:spatialdistr}
\end{figure*}

As a second test, we compared the spatial density distribution (Fig. \ref{fig:spatialdistr}) of our cluster, including stars within $r<4'$, with those of a sample of field stars selected at $5'\leq r \leq 8'$ from the cluster centre. As explained in our previous works \citep{ERG:submitted,Garro2021_19GCs},  we applied the Gaussian kernel density estimate (KDE; e.g.  \citealt{Rosenblatt1956,Parzen1962}) in order to better distinguish over-densities from lower densities than by using a visual inspection.  As expected, we find that the spatial distributions of the cluster and the field are very different since a central over-density is clear for the star sample centred on Garro~02 coordinates,  which decreases towards larger radii.\ Instead, no central over-density is visible for the field sample,  but a lumpy and non-homogeneous distribution is seen for it.  However, from the contour figures, the maximum number density of the cluster region is about six times the minimum level, which is not very different from the referenced field. Therefore, we explored the possibility of having statistical fluctuations by performing a third test. We constructed the radial density profile (RDP) for both cluster and stellar field samples. As expected, we find a flat radial density distribution for a stellar field, whereas a \cite{King1962} profile for the cluster sample. We postpone a detailed explanation of this test to Sections \ref{Decontproc} and \ref{sec:rdp}.\\

A fourth test regards both the comparison of vector proper motion (VPM) diagrams and also colour-magnitude diagrams (CMDs) for the cluster and star field samples. The former gives us information about the dynamics, whereas the latter allows the photometry of these two samples to be studied. Tests allow the main differences between the cluster and field samples to be shown. We postpone a detailed explanation of this test to Section \ref{Decontproc}.\\

Finally, these tests support the existence of an over-density of RGB stars, which could be associated with a new GC candidate.  Even so, a spectroscopic follow-up will confirm the real nature of Garro 02. We performed a photometric analysis, as explained in Sections \ref{Decontproc} and \ref{Parameters}.

\section{Datasets}
\label{datasets} 
We studied Garro 02 using cross-matched near-IR and optical datasets, adopting a $0.5''$ matching radius.  In particular, we used the near-IR dataset from the VVVX survey, acquired with the VISTA InfraRed CAMera (VIRCAM) at the 4.1m wide-field Visible and Infrared Survey Telescope for Astronomy (VISTA; \citealt{Emerson2010}).  The point spread function (PSF) photometry was extracted as described by \cite{AlonsoGarcia2018}. The astrometry was calibrated to the Gaia Data Release 2 (DR2; \citealt{Brown2018}) astrometric reference system, whereas the photometry was calibrated into the VISTA magnitude system \citep{Gonzalez_Fernandez2018} against the 2MASS using a globally optimised model of frame-by-frame zero points plus an illumination correction. The advantage of using this public survey is due to the fact that Garro~02 is located in an obscured region, which may be affected by differential reddening and high stellar crowding. Therefore, VVVX can help to see through the interstellar dust since it becomes more transparent at these wavelengths. Moreover, VVVX is capable of resolving individual stars better than other surveys, such as 2MASS \citep{Hajdu2020}.  Additionally,  the VVVX passbands ($JHK_s$) allow one to study evolved stars in the CMDs, such as red clump (RC) stars, which can be used to derive reliable distance estimates. \\

We also included data from 2MASS, a near-IR survey centred on the $JHK$ bands.  We utilized these data to build more complete CMDs since the brightest stars ($K_s <11$ mag) in the VVV images are saturated \citep{Saito2012}. We treated the 2MASS and VVVX data separately.  Therefore, we transformed the 2MASS photometry into the VISTA magnitude scale \citep{Gonzalez_Fernandez2018},  since the magnitude scale is different in the two photometric systems, noting very small differences of $\Delta J K_s < 0.03$ mag.\\

We also detected this cluster in the Gaia Early Data Release~3 (EDR3; \citealt{Brown2021}) and in the Pan-STARRS datasets (Figs. \ref{fig:tile} and \ref{fig:finderchart}).  However, we notice that the sampling of the area around Garro~02 is more homogeneous for the Gaia dataset than for Pan-STARRS. Indeed, we find that a greater number of stars are resolved in the Gaia survey; they also have more precise and reliable parameters (astrometry, PMs, and magnitudes).  Therefore, we preferred using only the Gaia EDR3 dataset for our analysis in order to benefit from its precise astrometry and exquisite PMs.  We excluded all nearby foreground stars with parallax $Plx>0.5$ mas, which is equivalent to $D<3$ kpc \citep{BailerJones2018}. We also used Gaia EDR3 PMs in order to perform the PM-decontamination procedure (Section \ref{Decontproc}).  Additionally, we made a cut using those stars with renormalised unit weight error \texttt{ruwe}<1.4 (\citealt{Fabricius2021}, \citealt{Brown2021}, and references therein), guaranteeing the high quality of
the astrometric solutions. No photometric colour or magnitude cuts were applied to the Gaia data.

\section{Decontamination procedure and cluster membership}
\label{Decontproc}
When studying the inner regions of the MW, various aspects have to be taken into consideration. First, the differential reddening and extinction, which not only affects the photometric distance estimates, but also the derived cluster age and metallicity, especially when isochrones are fitted.  Also, foreground and background field contamination can affect final results because it can both alter the total luminosity estimate and increase the errors in the cluster size derivation. 
However, we have used different procedures to overcome those problems and reduce uncertainties. \\

\begin{figure*}
\centering
\includegraphics[width=18cm, height=5.7cm]{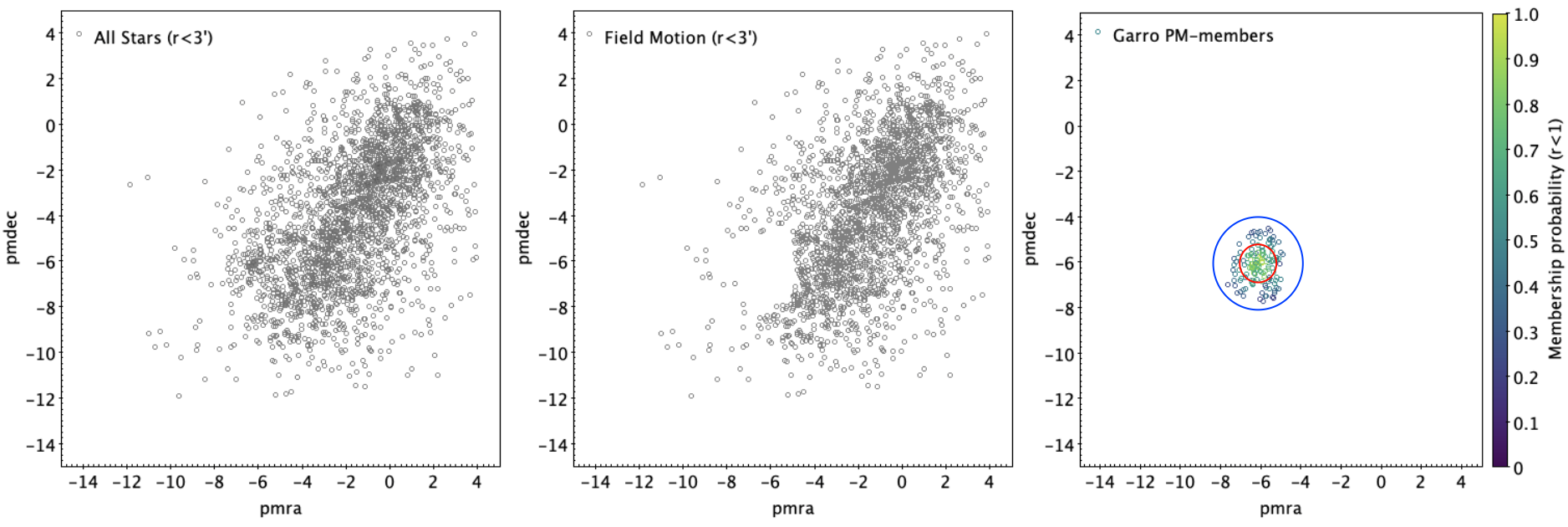}
\includegraphics[width=5.7cm, height=5.7cm]{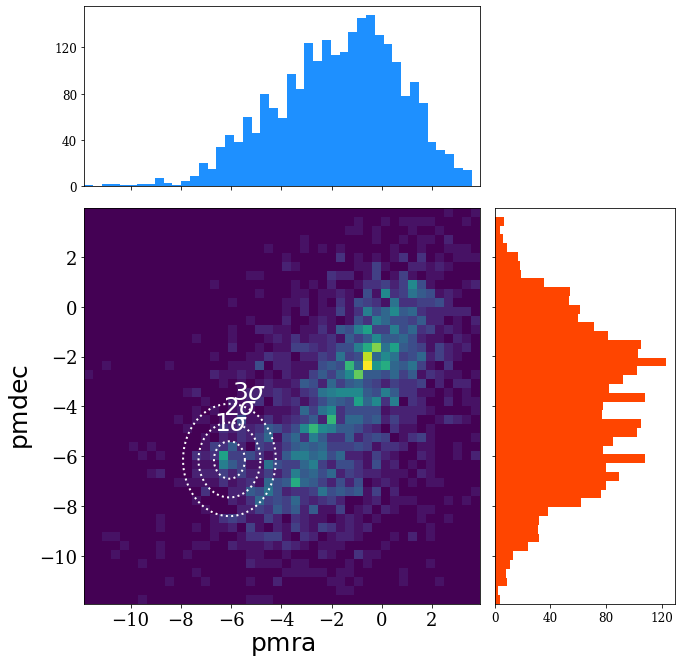} 
\includegraphics[width=6.1cm, height=5.7cm]{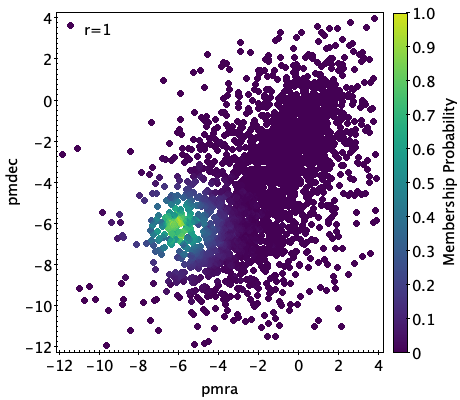} 
\includegraphics[width=6.1cm, height=5.7cm]{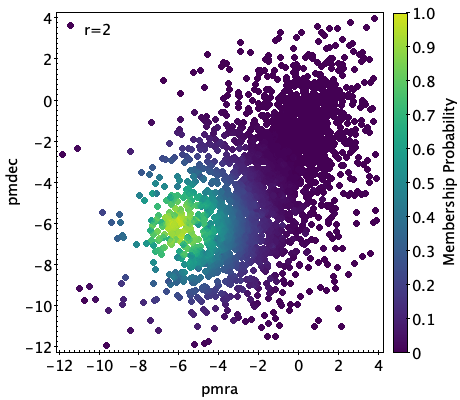} 
\caption{\textbf{Top}. VPM diagram for all star sample (left panel), field sample (middle panel), and Garro 02 PM members (right panel). In the right panel, the blue and red circles indicate a radius within 2 and 1 mas yr$^{-1}$, respectively.  As it is possible to see, we selected the most probable members ($>50\%$) within 1 mas yr$^{-1}$ from the mean cluster PMs.
\textbf{Bottom.} \textit{Left panel:} VPM diagram for Garro 02 cluster as a 2D histogram, displaying both the PMRA (in blue) and PMDEC (in red) histograms separately.  The yellow and light green colours depict over-densities, whereas the dark green and blue colours represent lower densities.  We indicate with dotted white circles the 1$\sigma$, 2$\sigma$, and 3$\sigma$ centred on the mean cluster PMs.  \textit{Middle and right panels:} VPM diagrams, where the brightest cloud represents the highest PM probability, within a radius  $r< 1$ mas~yr$^{-1}$ and $r< 2$ mas yr$^{-1}$.}
\label{fig:VPMs}
\end{figure*}

\begin{figure*}
\centering
\includegraphics[width=12cm, height=10cm]{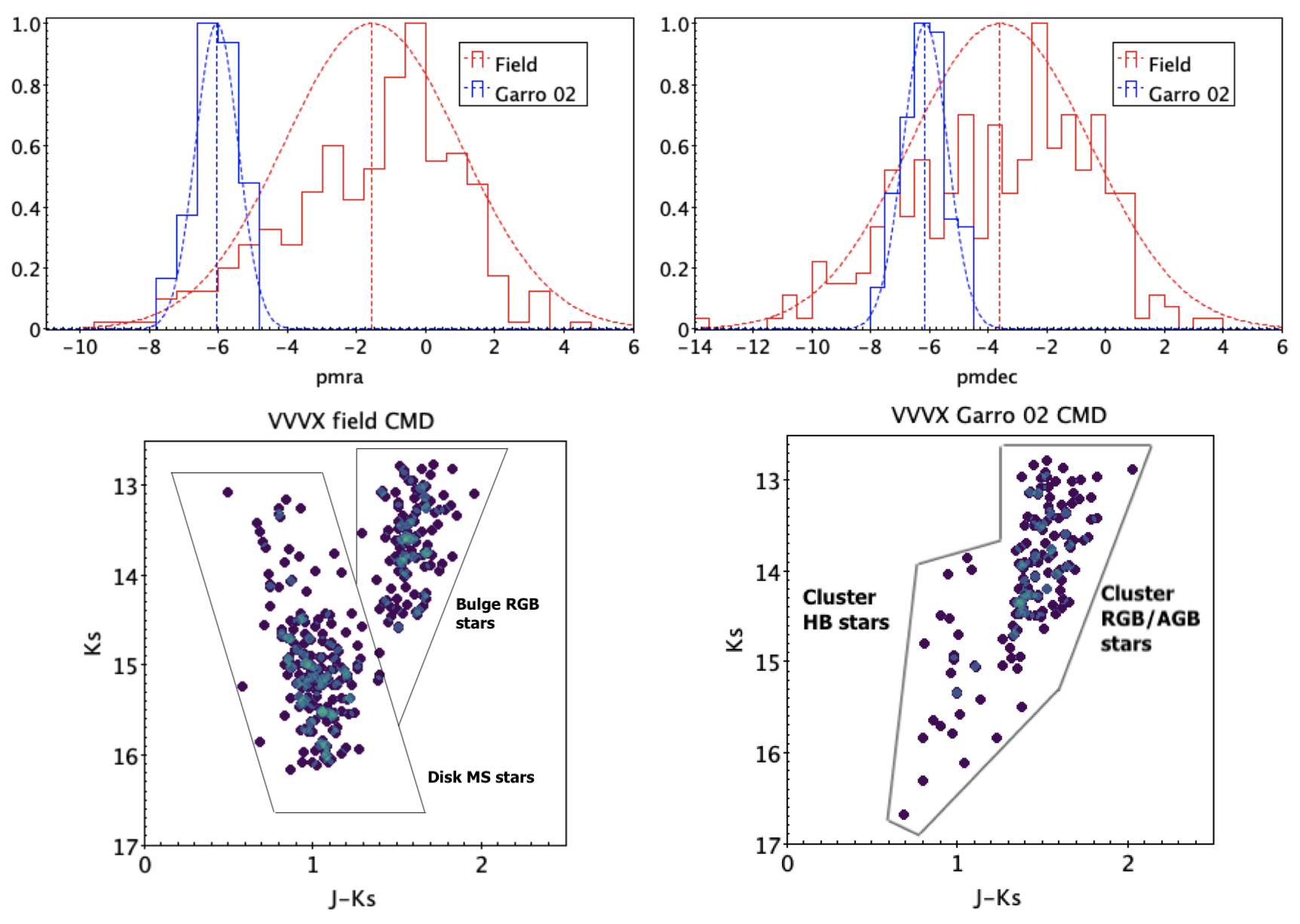} 
\caption{\textit{Top panels:} Normalised $\mu_{\alpha}^{\ast}$ (left) and $\mu_{\delta}$ (right) histograms for the field sample (in red) and Garro~02 cluster (in blue). We fitted the symmetric Gaussian distributions to highlight the differences in the peaks and in the dispersions of the two distributions.  \textit{Bottom panels:} Near-IR VVVX CMDs for the field sample (on the left) and for Garro 02 (on the right).  The green-coloured areas are representative of over-densities, while the blue ones are lower densities. }
\label{fig:comparisonIII}
\end{figure*}

First, we performed a decontamination procedure which allowed us to build clean CMDs and construct final catalogues with probable star cluster members. The decontamination of this dataset follows the one described by \cite{Garro2020,ERG:submitted,Garro2021_P99,Garro2021_19GCs}.  As mentioned in Section \ref{datasets}, we excluded all nearby stars, through a cut in parallax.  Thereafter, we included all stars within a given radius from the cluster centre.  We evaluated the Garro 02 dimension, using both a visual inspection of the density diagram as a function of the sky position and  applying the Gaussian KDE, as shown in Fig. \ref{fig:spatialdistr}.  Therefore,  we used the density map and the KDE iso-density contours in order to identify the more suitable cluster size, selecting the higher density area around the Garro 02 centre we adopted a $r<5.4'$. Finally, we preferred using the PM-decontamination procedure for two reasons: \textit{(i)} PMs ensure the stars in the cluster are divided better from the stars in the field since a cluster is expected to share coeval motions; \textit{(ii)} as demonstrated by \cite{Garro2021_P99} and \cite{Minniti2021_M48},  this method substantially reduces the percentage of contaminants better than the statistical decontamination procedure, excluding stars with disk-like PMs.  For this purpose,  we constructed the VPM diagrams as a 2D histogram shown in Fig. \ref{fig:VPMs}, in order to better identify the cluster PM peak. Indeed, as we can see in the bottom-left panel of Fig.\ref{fig:VPMs},  a highly peaked distribution is concentrated at $\mu_{\alpha}^{\ast}\approx -0.5$ mas yr$^{-1}$ and $\mu_{\delta}\approx -2.0$ mas yr$^{-1}$; while a detached and lower peaked distribution is located at $\mu_{\alpha}^{\ast} = -6.07\pm 0.62$ mas yr$^{-1}$ and $\mu_{\delta}= -6.15\pm 0.75$ mas yr$^{-1}$. The first is related to bulge+disk field stars, while the second one indicates the presence of the cluster. We derived the mean cluster PM (and its error) using the $\sigma$-clipping technique.  There is one concern that the PM dispersion is a bit large for a normal GC, but this may be due to the combination of different effects: the astrometric error, the residual contamination by field stars, or the possible ongoing GC disruption.  Also, we used this value in order to select stars within 1 mas yr$^{-1}$ from the mean cluster PM (see, top panels of Figure \ref{fig:VPMs}). 
Additionally, we calculated the PM membership probability (Fig. \ref{fig:VPMs}),  using a radius  $r< 1$ mas~yr$^{-1}$ and $r< 2$ mas yr$^{-1}$, in order to exclude as many contaminants as possible.  We can note from Figure \ref{fig:VPMs} that when we adopted $r<2$ mas yr$^{-1}$, we surely included a higher number of bulge+disk stars, hence we preferred adopting $r<1$ mas yr$^{-1}$. This allowed us to select only stars that are likely members of the cluster and thus  construct the final catalogue\footnote{The final catalogue for Garro 02 GC is available in upon request.  Please, write to elisaritagarro1@gmail.com.} including stars with $>50\%$ probability, subsequently minimising the likelihood of embedding contaminants. Even so, we expect that a very small fraction of contaminants may be present in the final catalogues due to the fact that these regions are characterised by high stellar crowding and by disk stars with similar PMs, as demonstrated by \cite{Garro2021_P99}.  In fact, as we can see from Figure \ref{fig:comparisonIII}, the two Gaussians (cluster and field samples) in the VPM diagram overlap.\  We have reduced the percentage of contaminants including only stars with a membership probability $>50 \%$; however, we estimate that $<6\%$ of contaminants persist in our catalogue after the cleaning.

Once the final catalogue is obtained,  we can build the CMDs for our cluster.  However, before performing the photometric analysis, we carried out a third and a fourth test,  mentioned in Section \ref{Discovery}.  In this case, we compare the VPM diagrams and the CMDs for both the cluster (using the decontaminated catalogue) and the field samples.  As shown in Fig. \ref{fig:comparisonIII},  we built the normalised PM histograms and we fitedt the symmetric Gaussian distributions, which show two very separated peaks ($\mu$) and very different dispersions ($\sigma$).  We find for Garro 02 cluster: $\mu^{Garro02}_{pmra}=-6.07$  mas yr$^{-1}$ with $\sigma^{Garro02}_{pmra}= 0.62$  mas yr$^{-1}$, and $\mu^{Garro02}_{pmdec}=-6.15$  mas yr$^{-1}$ with $\sigma^{Garro02}_{pmdec}= 0.75$  mas yr$^{-1}$. Whereas, for the field sample: $\mu^{field}_{pmra}=-1.57$  mas yr$^{-1}$ with $\sigma^{field}_{pmra}= 2.56$ mas yr$^{-1}$, and $\mu^{field}_{pmdec}=-3.60$  mas yr$^{-1}$ with $\sigma^{field}_{pmdec}= 3.07$  mas yr$^{-1}$.\\
However, it is notable that a symmetric Gaussian distribution is a rough fit for the field sample, but this is expected since the field stars show some structured motions, but their broad and skewed PM distributions are very different from the cluster that is more concentrated and nearly Gaussian, which is as expected for a GC.\\
Finally, we used the CMDs as excellent tools to understand the cluster nature.  In fact, comparing the cluster and field near-IR VVVX CMDs (Fig. \ref{fig:comparisonIII}), we can see that the blue horizontal branch (HB) is visible in the cluster CMD, suggesting a low metallicity, and also a narrow red giant branch (RGB) sequence is evident. On the other hand, no clear sequences are displayed by the field CMD, but a very elongated blue sequence is shown, indicating an excess in the disk stars' main sequence, as well as wide red sequence,  belonging to the RGB stars from the bulge. Unfortunately,  fainter stars belonging to the cluster MS turn off are missing in our compilation since they just are below the magnitude detection limit.\\

\begin{figure}[htb]
\centering
\includegraphics[width=8cm, height=5cm]{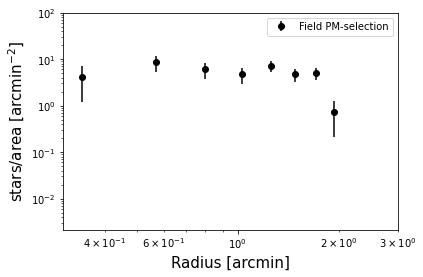} 
\caption{Flat radial density profile for the PM-selected field stars. The black points correspond to the field density profile obtained with no subtraction of the background contribution. }
\label{fig:rdp_field}
\end{figure}

The last test we carried out was to probe the presence or lack thereof of statistical fluctuations in the Garro 02 regions is comparing the RDP of Garro~02 with the field sample. We have dedicated a section (see Sect. \ref{sec:rdp}), where we explain how we constructed the RDP for our target. Here, we just show the RDP for the field sample. We preferred using a sub-sample of stars, selecting only those within 1.5 mas yr$^{-1}$ of the mean stellar field PMs, which were obtained previously, in order to make a sensible comparison between the two RDPs.  Figure \ref{fig:rdp_field} shows the RDP for the PM-selected field sample. As expected for these areas, we find a flat distribution, with a constant value of 5-8 stars arcmin$^{-2}$, while as explained in Section \ref{sec:rdp},  a King profile represents a good approximation for the Garro~02 RDP.
Finally, these features confirm the cluster nature of Garro~02.

\section{Photometric characterisation of Garro 02}
\label{Parameters}
Once decontaminated catalogues with probable members are built, we can measure the main physical parameters that characterise the cluster, such as the reddening and extinction, the distance modulus and heliocentric distance, the metallicity and age, the total luminosity, and also the main structural parameters, such as the tidal radius ($r_t$) and core radius ($r_ c$).

\subsection{Reddening, extinction, and distance determinations}
\label{redd_ext_dist}
Although the advantage of using multi-wavelength photometry, as in our case, is that it can help to reduce the sources of error substantially and reach robust results, it is often tricky to find all the parameters which fit both optical and near-IR CMDs simultaneously.  For that reason, we derived the reddenings and extinctions using different methods. 

First, we used the reddening maps of \cite{Schlafly2011}, recovering an optical extinction of $A_{V}=8.57\pm 0.02$ mag.  Therefore,  adopting the known extinction relation $A_{Ks}~=~0.11~\times~A_{V}$ \citep{Schlafly2011}, we find in the near-IR $A_{Ks}= 0.94\pm 0.02 $ mag.  On the other hand,  we used two extinction coefficients for deriving the excess colour: $R_{Ks}=0.428$ \citep{AlonsoGarcia2017} was obtained for the inner bulge and $R_{Ks}=0.75$ \citep{Cardelli1989} was mostly used for regions in the outer bulge; they allowed us to calculate an excess colour of $E(J-Ks)=2.20\pm 0.05 $ mag and $E(J-Ks)=1.27\pm 0.03$ mag, respectively.  In the first case, we measured a distance modulus of $(m-M)=14.16\pm 0.05 $ mag, equivalent to $D=6.8 \pm 0.2 $ kpc, whereas in the second case the distance modulus is $(m-M)=13.64\pm 0.03 $ mag, which is equivalent to a heliocentric distance of $D=5.4\pm 0.2$ kpc.

\begin{figure}[htb]
\centering
\includegraphics[width=6cm, height=6cm]{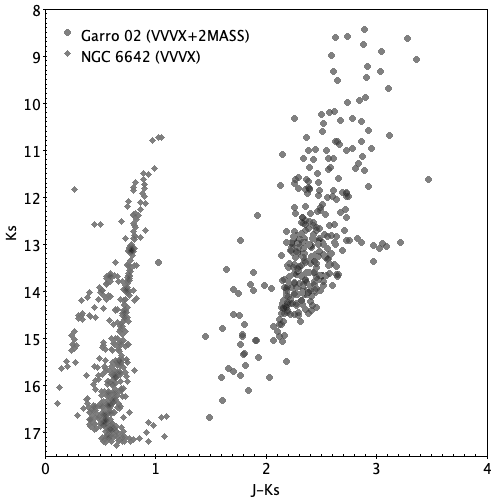} 
\caption{Comparison between the near-IR CMDs for Garro 02 (grey dots) and NGC 6642 (grey diamonds). We shifted Garro 02 by 0.8 in $(J-K_s)$ colour to avoid overlapping points. }
\label{fig:Garro02+NGC6642}
\end{figure}

Second, we built near-IR CMDs via comparison with the well-known Galactic GC NGC 6642, which is located at a distance of $D=8.1\pm 0.2$ kpc with a distance modulus of $(m-M)=14.53\pm 0.17$ mag \citep{Balbinot2009,BaumgardtVasiliev2021}. We compared these two clusters (see Fig. \ref{fig:Garro02+NGC6642}), because they show similar RGB and HB morphologies. Clearly, we can see a complete CMD, showing very narrow and well-defined sequences for NGC 6642.  Whereas, for Garro 02, we can see a wider RGB due to a higher reddening in this field,  as well as a scarcely populated HB due to the fact that Garro 02 is fainter than NGC 6642. Indeed, we observed that their sequences are separated by $\Delta(J-K_s)=0.86$ mag,  with Garro 02 being more reddened, and $\Delta(K_s)= 0.09$, with Garro 02 being slightly fainter. Taking into account that the NGC 6642 reddening is $E(J-K_s)=0.21\pm 0.06$ mag, this yields a reddening $E(J-K_s)=1.07\pm 0.06$ mag. The differential comparison with NGC 6642 implies a distance modulus of $(m - M) = 13.79 \pm 0.06$ mag, equivalent to $D = 5.7\pm 0.7$ kpc.\\

Independently,  we derived reddening and extinction in the optical passbands.  Since clusters with blue HBs usually do not show a defined RC, as Garro 02, we derived its colour excess, comparing the cluster HB and RGB colours with the absolute colours by \cite{Babusiaux2018}. We obtained a reddening of $E(BP-RP) = 2.40\pm 0.01$ mag and an extinction $A_{G} = 4.80\pm 0.02 $ mag, from which we calculated a distance modulus of $(m-M)_0 =13.32\pm 0.02$ mag, which is equivalent to a heliocentric distance of $D=4.6\pm 0.2$ kpc.  Knowing that  $A_{G} = 10.116 \times A_{Ks}$, we derived the near-IR extinction  $A_{Ks} = 0.47\pm 0.01$ mag, which allowed us to obtain a distance modulus of $(m-M)_0 = 14.07\pm 0.01$ mag, equivalent to $D=6.5 \pm 0.2$ kpc. \\
Hence, we find that all these distance estimates are all in agreement within errors ($<2\sigma$). We adopted a mean value of $D=5.6\pm 0.8$ kpc, as the cluster distance, thus we placed Garro 02 at a Galactocentric distance of $R_{G}=2.9$ kpc, assuming $R_{\odot}=8.2$ kpc \citep{Gravity2019} and at a distance above the Galactic plane of $Z=0.006$ kpc,  using the relation $Z=D\times \sin (b)$ and assuming $Z_{\odot}=0$ kpc. Finally, once distance and mean cluster PMs were evaluated, we could derive the tangential velocities, finding $V_{T}^{RA}=-163.87\pm 33.62$ km s$^{-1}$ and $V_{T}^{Dec}=-166.26\pm 16.15$ km s$^{-1}$. Therefore, the position and motions indicate that Garro 02 is a new bulge cluster.

\subsection{Metallicity and age estimates}
\label{subs:metaage}
\begin{figure*}
\centering
\includegraphics[width=10cm, height=6cm]{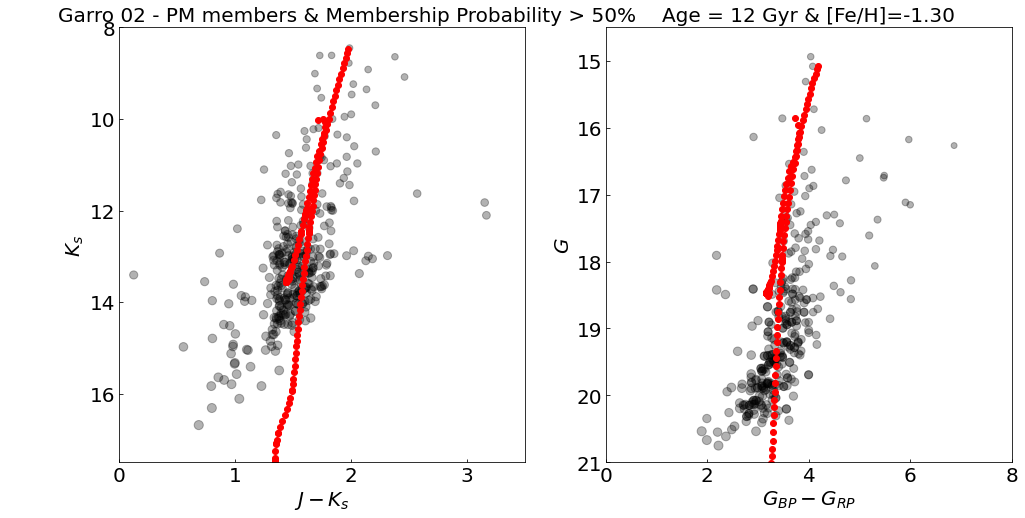} 
\caption{Optical Gaia EDR3 (right panel) and near-IR VVVX+2MASS (left panel) CMDs for the GC Garro 02.  We have included only PM-selected members, considering also those stars with a membership probability greater than 50\%.  The isochrone-fitting method yields an Age = 12 Gyr and [Fe/H] = -1.30 dex. We have superimposed the PARSEC model represented as red dotted lines in both panels. }
\label{fig:CMDs}
\end{figure*}

\begin{figure*}
\centering
\includegraphics[width=4cm, height=4cm]{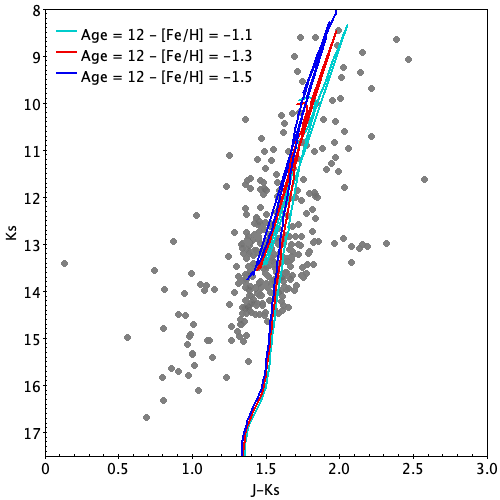} 
\includegraphics[width=4cm, height=4cm]{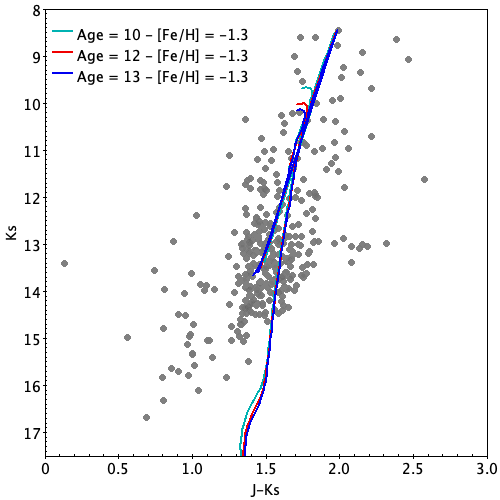} 
\includegraphics[width=4cm, height=4cm]{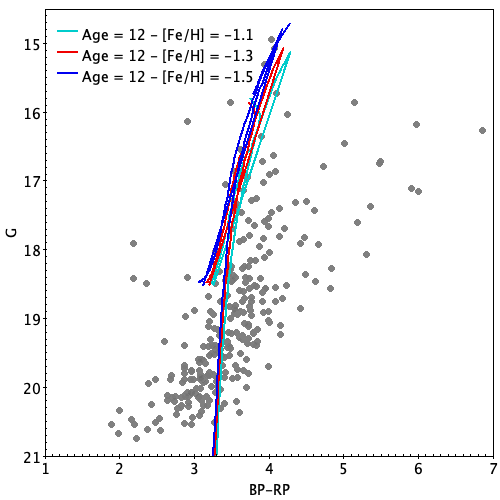} 
\includegraphics[width=4cm, height=4cm]{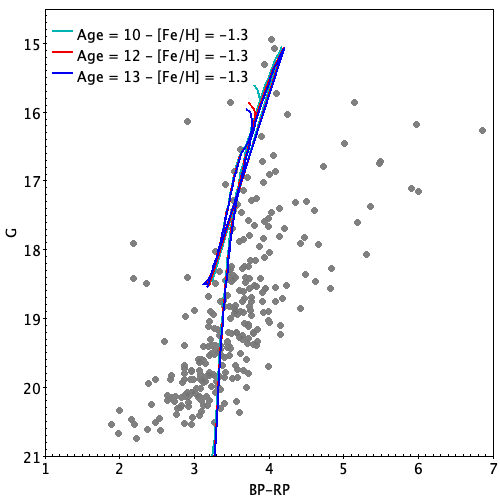} 
\caption{Optical Gaia EDR3 (middle and right panels) and near-IR VVVX+2MASS (middle and left panels) CMDs for the GC Garro 02.  We have included only PM-selected members, considering also those stars with a membership probability greater than 50\%.  We used a family of isochrones, changing metallicity and age in turn,  in order to limit the age-metallicity degeneracy and derive their errors. For our convenience, we specify here that the unit of the age is `Gyr' and that of the metallicity is `dex' in the legend.}
\label{fig:CMDs2}
\end{figure*}

In the previous section (Sect. \ref{redd_ext_dist}),  we derived the reddening and the extinction with different methods. Therefore, we used these values in order to fit a family of isochrones in the near-IR and optical CMDs. We found that the parameters that allow us to better reproduce the evolutionary sequences in the CMDs are those reported in Table \ref{table1}, which were derived using the comparison with the GC NGC 6642. Once these parameters were fixed, we performed the isochrone-fitting methods in order to derive the metallicity and age of our target.  From the CMD morphology of this cluster and the similarities with the GC NGC 6642 CMD, we expect that Garro 02 is a metal-poor and old cluster.  \\
We downloaded the PARSEC+COLIBRI isochrones\footnote{http://stev.oapd.inaf.it/cgi-bin/cmd}  \citep{Bressan2012,Marigo2013}.  
We first generated a family of isochrones, considering a metallicity range of  $-1.0\leqslant$ [Fe/H] $\leqslant -2.0$ dex,  and we superimposed these tracks on the CMDs both in near-IR and optical passbands (Fig. \ref{fig:CMDs}).  After that, we followed the same procedure with ages,  generating isochrones with ages between 8 and 13 Gyr.  Hence, comparing the evolutionary sequences in the Garro 02 CMDs with the isochrones generated for different ages and metallicities, we visually selected the best fit, yielding a [Fe/H]~=~$-1.30\pm 0.2$ dex and Age $=12 \pm 2$ Gyr.  We provide an estimation of the age and metallicity errors by changing the parameters until the isochrones are not adequate to reproduce all evolutionary sequences both in Gaia and VVVX+2MASS CMDs.  Recovering the metallicity is simpler than the age since the more evolved sequences, such as the HB and the RGB, are more affected by variations in metallicity,   whereas deriving a precise age estimate is a tricky task to perform,  especially when stars fainter than the main sequence turn-off (MSTO) are below the detection limit (see Figure \ref{fig:CMDs2}).  Despite this, the vertical (as well as the horizontal) method (e.g. \citealt{Gratton1987,Sarajedini1989,MarinFranc2010}) is largely used for the relative age determination. The estimate of the magnitude difference between the TO and the HB level is a good relative age indicator for a cluster. The advantage of using this method is that it is independent of the reddening, distance modulus, and metallicity.  We can see from Figures \ref{fig:CMDs} and \ref{fig:CMDs2} that the difference in magnitude between the HB and MSTO is $\Delta K_s(HB-MSTO)>3$ mag and $\Delta G(HB-MSTO)>2.5$ mag, indicating that our cluster is older than 10 Gyr.  This method helped us to limit the age-metallicity degeneracy, since we know that ages < 10 Gyr are not suitable for Garro 02. However, this represents a lower limit since our CMDs do not reach the MSTO.  From our analysis, we can conclude that Garro 02 is a metal-poor and old bulge GC.

\subsection{Total luminosity derivation}
As mentioned in Section \ref{subs:metaage},  MS faint stars are missing in our compilation; therefore, it is complicated to derive a precise estimate of the total luminosity for this GC. However,  as demonstrated in our previous works \citep{ERG:submitted,Garro2021_SGR,Garro2021_P99,Garro2021_19GCs}, we can derive a good estimation of this parameter, comparing the luminosity of Garro 02 with those of known and well-characterised GCs, as long as they have a similar metallicity\footnote{We consider the metallicity as good parameter for comparing GCs because it has an effect, as seen in theoretical isochrones for example. In general, metal-poor GCs should be bluer, and therefore brighter, than metal-rich GCs at the same mass.}.  We first derived the absolute magnitude, measuring the cluster total flux. We obtain $M_{Ks}= -7.52\pm 1.23$ mag, which is equivalent to $M_{V}=-5.24$ mag, if we assume the typical GC colour of $(V-K_s)=2.5$ (e.g. \citealt{Barmby_2000,Conroy_2010}). Then, we applied the procedure described by \cite{Garro2021_SGR} in order to quantify the fraction of luminosity missed.  In summary, we first calculated the absolute magnitude in $K_s$-band for the Galactic GCs, using the VVV/VVVX catalogues, thus in the same magnitude ranges as Garro~02 ($10.2 \lesssim K_s \lesssim 17.2$ mag). After that,  the obtained $M_{Ks}$ was converted into the absolute magnitude in V-band $M_V$, using again the typical GC colour.  Finally,  we scaled the $M_V$ values to the absolute magnitude listed in the 2010 version of the Harris (1996) catalogue,  estimating the fraction of luminosity that comes from the faintest stars for the known GCs.  In particular,  we compared Garro 02 with NGC 6642 ([Fe/H] $=-1.26$, \citealt{Barbuy2006}) and NGC 6626 ([Fe/H] $=-1.32$, \citealt{Kerber_2018}). Therefore, we calculated the average of $\Delta(M_V)$ fractions for these two clusters and finally we added this average to the Garro~02 luminosity.  We estimate that this fraction of missed luminosity is $\sim 0.4$ mag (on average), and therefore fainter stars do not strongly change the total luminosity.  In this way, we have obtained an empirical correction to our cluster's  luminosity under the assumption of similarity among Galactic GCs.  We obtain a total luminosity in V-band of $M_V = -5.44$ mag,  thus finding a low luminosity GC, that falls in the fainter tail of the MW GC luminosity function \citep{Rejkuba2012}. It is notable that the total luminosity should also be uncertain because the luminosity function of a cluster may be altered within a strong tidal field. However, as demonstrated by \cite{Kharchenko2016}, we expect that the first 10–12 brighter members accumulate more than half of the integrated luminosity of a cluster;  therefore, our luminosity estimate is a lower limit of  the real total luminosity of the GC since both optical and near-IR catalogues are incomplete.

\subsection{Radial density profile}
\label{sec:rdp}
We computed the RDP for Garro 02 with the purpose of determining its true physical size.
The old centre was initially found by visual inspection of the available near-IR images and density maps, and it should therefore be more uncertain than the new centre determination.  Therefore, we checked that the centre coordinates of the system were accurately determined for a proper determination of the RDP. Various papers, presented in the literature,  explain different methods for deriving the cluster centre, some of them are based on the stellar counts (e.g. \citealt{Lanzoni2019}) or the position of the maximum of the surface brightness \citep{Trager1995}.  Since our compilation includes only evolved stars, we used the maximum spatial density determined through a 2D Gaussian KDE density estimator in order to assess the location among our points (or stars) where the maximum density is found.  In this way we can avoid any possible bias induced by the presence of a few bright stars, which would alter the position of the surface brightness maximum.  \\
We first divided the distribution of points in a grid both in RA and Dec, between 0$^\circ$ (coincident with the cluster centre) and 0.08$^\circ$, with a step of 0.005$^\circ$. For estimating the point density using a kernel, we set two functions. The first one performs two tasks at once: it takes two points defined by RA1, DEC1, RA2, and DEC2; calculates their Euclidean distance; and uses that distance to evaluate a Gauss kernel's function value. The second function transforms the Gauss kernel to approximately yield $1$ when the distance is $0$ (or very small), and to have 2$\sigma$ fixed at the previously calculated radius. After that, the code identifies the maximum value in each grid and gives us the maximum of the density distribution of our cluster. In this way, we find that the new centre is $RA =$  18:05:54.0 (equivalent to $271.475^{\circ}$) and $Dec=$ -17:42:00 (equivalent to $-17.700^{\circ}$) in the equatorial system, and $l=12.048^{\circ}$ and $b=1.646^{\circ}$ in the galactic system, which slightly differ from the old centre of  $\Delta RA=-0.0121^{\circ}$ and $\Delta Dec = -0.0002^{\circ}$. \\

\begin{figure*}
\centering
\includegraphics[width=6.5cm, height=6cm]{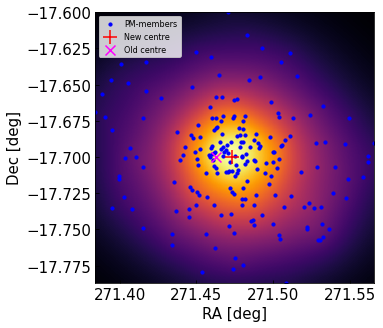} 
\includegraphics[width=8cm, height=5.5cm]{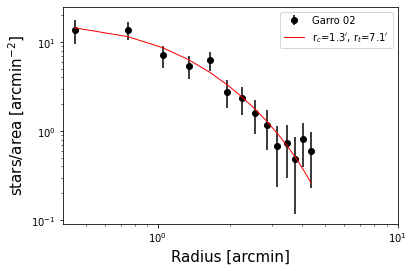} 
\caption{\textit{Left panel:} Determination of the Garro 02 centre using the KDE technique.  As specified by the legend, the blue points are the PM selected cluster members, while we highlight the old centre with a magenta cross, and the new centre with a red plus sign. The yellow colour depicts higher densities, which decreasing to lower densities becomes orange, blue, and black. \textit{Right panel:} The black points correspond to the cluster density profile obtained after subtraction of the background contribution.  The red line shows the best-fit King model profile, with the corresponding values of the core radius of $r_c =1.3'$ and a tidal radius of $r_t=7.1'$.  }
\label{fig:RDP}
\end{figure*}

We used this new centre in order to construct the RDP for our target. We divided our sample into 14 circular annuli, increasing radii of $0.3'$ out to an outer radius of $4.8'$.  Although our statistics are low because they are only related to evolved stars, this allowed us to include a good number of stars in each bin ($N\geqslant 7$ stars).  We calculated the density per bin as the number of stars (N) divided by the area (A). The errorbars were computed as $e = \sqrt{N}/A$. We also derived the true RDP correcting our profile subtracting the background level, fixed at 0.5 stars/arcmin$^{2}$.  Thereafter, the RDP for Garro 02 was plotted as function of the mean distance of the circular annulus to the cluster centre (see Fig. \ref{fig:RDP}).  \\
One of the most used models to fit the cluster density profile is the \cite{King1962} model. Even so,  the true shape should likely be more complex; since we base our study on evolved stars which may be more concentrated, we find that the \cite{King1962} model is a good representation of the projected number density. Also, it allows us to provide the cluster main structural parameters.  The best-fit King model yields a core radius of $r_c= 1.25\pm 0.27$ arcmin and a tidal radius of $r_t= 7.13\pm 3.83$ arcmin, which are equivalent to $r_c= 2.04\pm 0.44$ pc and $r_t= 11.61 \pm 6.24$ pc at the cluster distance of $D=5.6$ kpc.  These values are consistent with the typical MW GC radii listed in the 2010 \cite{harris1996catalog} compilation. However, owing to the low density and also no well-resolved profile due to the low statistics especially at smaller radii, we are not able to suggest if Garro 02 is a pre-, post- or core-collapse GC.  However, calculating the concentration parameter $c=\log(r_t/r_c) = 1.74$ and comparing this value with GCs listed in the 2010 \cite{harris1996catalog} compilation, we exclude that Garro~02 is a core-collapse cluster.

\section{Summary and conclusions }
\label{Conclusions}
We report the discovery of a new GC candidate located in the MW bulge, named Garro 02.  We found this cluster as a clear over-density of red giant stars in the VVVX density map (Fig. \ref{fig:tile}), and also the cluster PMs peak differ from the bulk of the field PMs (Fig. \ref{fig:VPMs}).  We confirmed the cluster nature performing four independent tests, making a comparison between the spatial distributions,  radial density distributions, VPM diagrams, and CMDs between the cluster and the field samples. Thereafter, we carried out a multi-band photometric analysis, using a combination of three surveys: VVVX and 2MASS in the near-IR, and Gaia EDR3 in the optical passbands.  We measured the cluster's physical parameters, including reddening, extinction, distance, metallicity, age,  total luminosity, and also structural parameters deriving core and tidal radii. Our main results for the parameters of Garro~02 are listed in Table \ref{table1}. \\ 
Garro 02 is a genuine GC located in the Galactic bulge at a heliocentric distance of $5.6$ kpc and a galactocentric distance of $2.9$ kpc. It is a metal-poor ([Fe/H] = $-1.30$) and old (age $\sim 12$ Gyr) GC. Additionally, it is a low-luminosity object with $M_{V}=-5.44$ mag,  which is fainter than the MW GC luminosity function peak ($M_V^{peak} = -7.4\pm 0.2$, \citealt{Harris_1991,ashman_1998}, and in agreement with the
value found by \citealt{Garro2021_SGR} of $M_{V}^{peak} = -7.46\pm 1.04$ mag), suggesting that this cluster may have survived a strong dynamical process.\\
Finally,  deeper observations need to be done in order to reach the main sequence to provide a more precise age estimate and complete the Garro 02-star catalogue, including fainter stars, to confirm its structural parameters and better calculate the total luminosity for this cluster.  Spectroscopic analysis needs to be performed in order to confirm the metallicity found by the present work, and also to derive radial velocities and chemical abundances. 

\begin{table}
\centering 
\caption{Main physical parameters for Garro 02 determined using near-IR and optical datasets.}
\begin{tabular}{lc}
\hline\hline
\textbf{Physical parameters} & \textbf{Garro 02} \\
\hline
RA (J2000) & 18h05m54.0s  \\
DEC (J2000) & -17d42m00s  \\
Latitude  & $ 12^{\circ}.048 $\\
Longitude   & $ +1^{\circ}.646 $\\
\hline
$A_{K_s}$ [mag]& $ 0.79\pm  0.04$ \\
$A_{G}$ [mag]&  $ 4.80 \pm 0.02$  \\
$E(J-K_{s})$ [mag]& $ 1.07\pm 0.06$  \\
$E(BP-RP)$ [mag]& $ 2.40\pm 0.01$  \\
$D_{mean}$ [kpc] &$ 5.6\pm 0.8 $\\
$R_{G}$ [kpc]& $2.9$\\
$Z$ [kpc] & $0.006$\\
\hline
$M_{K_s}$ [mag]& $-7.52\pm 1.23$ \\
$M_{V}$ [mag] & $-5.44$ \\
\hline
[Fe/H] [dex]& $ -1.30\pm0.2$ \\
Age [Gyr]&  $12.0\pm 2$  \\
\hline
$\mu_{\alpha}^{\ast}$ [mas/yr]& $-6.07 \pm 0.62 $ \\
$\mu_{\delta}$ [mas/yr]& $-6.15 \pm 0.75$ \\
$V_{T}^{RA}$ [km/s] & $-163.87\pm 33.62$\\
$V_{T}^{Dec}$ [km/s] & $-166.26\pm 16.15 $\\
\hline
$r_c$ [arcmin] & $1.25 \pm 0.27$ ($2.04$ pc)\\
$r_t$ [arcmin] & $7.13\pm 3.83$ ($11.61$ pc)\\
\hline\hline
\end{tabular}
\label{table1}
\end{table}

\begin{acknowledgements}
The authors are grateful to the anonymous referee for providing helpful comments and suggestions, which have improved the content of this paper. We gratefully acknowledge the use of data from the ESO Public Survey program IDs 179.B-2002 and 198.B-2004 taken with the VISTA telescope and data products from the Cambridge Astronomical Survey Unit. ERG acknowledges support from ANID PhD scholarship No. 21210330.  D.M. gratefully acknowledges support by the ANID BASAL projects ACE210002 and FB210003, and Fondecyt Project No. 1220724. This work has made use of data from the European Space Agency (ESA) mission Gaia (http:/www.cosmos.esa.int/gaia), processed by the Gaia Data Processing and Analysis Consortium (DPAC, http://www.cosmos.esa.int/web/gaia/dpac/consortium).  J.G.F-T gratefully acknowledges the grant support provided by Proyecto Fondecyt Iniciaci\'on No. 11220340, and also from ANID Concurso de Fomento a la Vinculaci\'on Internacional para Instituciones de Investigaci\'on Regionales (Modalidad corta duraci\'on) Proyecto No. FOVI210020, and from the Joint Committee ESO-Government of Chile 2021 (ORP 023/2021).  J.A.-G. acknowledges support from Fondecyt Regular 1201490 and from ANID – Millennium Science Initiative Program – ICN12\_009 awarded to the Millennium Institute of Astrophysics MAS.
\end{acknowledgements}

\bibliographystyle{aa}
\bibliography{bibliopaper6}

\end{document}